\documentclass[10pt]{article}
\usepackage{pack_manage}

\newif\ifshowabstract
\showabstracttrue

\newif\ifshowintroduction
\showintroductiontrue

\newif\ifshowmethodology
\showmethodologyfalse

\newif\ifshowMaterialsMethods
\showMaterialsMethodstrue

\newif\ifshowresults
\showresultstrue

\newif\ifshowDiscussion
\showDiscussiontrue

\newif\ifshowappendix
\showappendixfalse

\begin{document}
\fontsize{12pt}{14.4pt}\selectfont  

\title{\textbf{Tau-induced atrophy drives functional connectivity disruption in Alzheimer’s disease}}
\author[a]{Kun Jiang}
\author[b]{Can Liao}
\author[c]{Sujin Jiang}
\author[d]{Haidong Lin}
\author[a]{Jixin Hou}
\author[e]{Tianming Liu}
\author[f]{Gang Li}
\author[c]{Taotao Wu}
\author[g]{Yiqi Mao}
\author[h]{Ellen Kuhl}
\author[a,1]{Xianqiao Wang}
\author[i]{Xianyan Chen}
\affil[a]{School of ECAM, College of Engineering, University of Georgia, Athens, GA, 30602, USA}
\affil[b]{Institute of Bioinformatics, University of Georgia, Athens, GA, 30602, USA}
\affil[c]{School of Chemical, Materials, and Biomedical Engineering, College of Engineering, University of Georgia, Athens, GA 30602, USA}
\affil[d]{Department of Engineering Mechanics, Zhejiang University, Hangzhou, Zhejiang, 310027, China}
\affil[g]{Department of Engineering Mechanics, College of Mechanical and Vehicle Engineering, Hunan University, Changsha, Hunan, 410082, China}
\affil[i]{Department of Epidemiology and Biostatistics, College of Public Health, University of Georgia, Athens, GA 30602, USA}
\affil[e]{School of Computing, University of Georgia, Athens, GA 30602, USA}
\affil[f]{Department of Radiology and Biomedical Research Imaging Center, The University of North Carolina at Chapel Hill, NC 27599, USA}
\affil[h]{Department of Mechanical Engineering, Stanford University, Stanford, CA 94305, USA}\vspace{-1em}

\begingroup
\let\center\flushleft%
\let\endcenter\endflushleft%
\maketitle
\endgroup

\selectlanguage{english}

\begin{abstract}
\large
\ifshowabstract
    Alzheimer’s disease involves progressive tau accumulation and spread, leading to regional brain atrophy and disruption of large-scale functional networks. While tau propagation and tissue degeneration have been widely modeled, how atrophy dynamics translate into functional connectivity (FC) degradation remains unclear. Here, we develop a multiphysics framework integrating anisotropic tau reaction–diffusion, finite-deformation biomechanics, and network modeling to link tau-driven atrophy with FC changes. Model fidelity is evaluated by quantitatively comparing simulated atrophy patterns with imaging-derived measurements. Using longitudinal structural and functional MRI, we identify an approximately linear relationship between regional atrophy rates and FC change. We then construct an atrophy-informed structural network degradation matrix from model-predicted region-specific atrophy rates and embed it into a neural oscillation model to predict FC disruption. Our results show that (i) the coupled reaction–diffusion–biomechanical model reproduces observed regional atrophy, (ii) regional atrophy rates parsimoniously predict longitudinal FC changes, and (iii) the atrophy-informed degradation matrix captures the direction and relative magnitude of regional FC disruption. By converting tau-driven atrophy into predictive FC trajectories, the proposed framework offers a clinically interpretable avenue for forecasting disease progression and informing trial design.

\fi
\end{abstract}%

\sloppy
\newpage
\section{Introduction}
\ifshowintroduction
    The accumulation of pathological amyloid-$\beta$ and hyperphosphorylated tau protein is a classical hallmark of Alzheimer's disease, emerging years to decades before a clinical diagnosis becomes possible \citep{duyckaerts_classification_2009}. The widely accepted amyloid cascade hypothesis posits amyloid-$\beta$ as the initiating factor, triggering a sequence of downstream events, including tau misfolding and propagation, which subsequently lead to region-specific brain atrophy and ultimately cognitive impairment \citep{jack_biomarker_2013}. In parallel with growing availability of multimodal imaging, recent work has sought to clarify the mechanisms of tau propagation and its downstream effects by employing a broad range of approaches such as physical modeling and statistical analyses \citep{weickenmeier_multiphysics_2018, schafer_interplay_2019}. However, despite these advances, the quantitative relationship between atrophy progression and subsequent cognitive impairment remains insufficiently characterized. This gap is particularly consequential given that Alzheimer’s disease evolves along a prolonged subclinical trajectory, with biomarker and neurodegenerative changes accumulating for years before overt symptoms emerge (often spanning 7--11 years)\citep{raket_estimating_2025}. Establishing a mechanistic and predictive link between atrophy dynamics and cognitive decline is therefore essential for improving the diagnosis and prognosis of Alzheimer’s disease, particularly in addressing clinically relevant questions such as: \textit{How rapidly will cognitive function deteriorate}? 

Histological evidence indicates that the progression of Alzheimer’s disease is closely associated with the spatiotemporal spreading of misfolded protein aggregates throughout the brain \citep{braak_neuropathological_1991}. In the brains of patients with Alzheimer’s disease, substantial accumulation of amyloid-$\beta$ and tau protein has been observed, with tau pathology being recognized as the primary driver of brain tissue atrophy. Tau deposition is first detected in the entorhinal cortex and the locus coeruleus, and subsequently advances to anatomically and functionally interconnected cortical regions \citep{braak_neuropathological_1991}. Due to the intracellular transport mechanisms of tau protein, axonal transport plays a critical role in the dissemination of misfolded protein aggregates across neural pathways \citep{jucker_pathogenic_2011,jacobs_structural_2018}. Recent studies have further revealed that the spatiotemporal patterns of neurodegeneration exhibit striking similarities to prion-like propagation processes, suggesting a self-propagating mechanism underlying disease progression \citep{jucker_self-propagation_2013}. Motivated by these observations, numerous studies have employed nonlinear diffusion models based on the Fisher–Kolmogorov equation to reproduce protein spreading in various neurodegenerative diseases, achieving strong agreement with experimental and clinical data \citep{weickenmeier_multiphysics_2018, fornari_prion-like_2019}. By leveraging multiphysics finite element frameworks, tau protein propagation can be explicitly coupled with tissue atrophy, enabling quantitative investigation of their mutual interactions. The predicted spatiotemporal patterns of protein spreading and brain atrophy show good consistency with clinical observations inferred from in vivo magnetic resonance imaging as well as post-mortem histopathology \citep{schafer_interplay_2019, pederzoli_coupled_2024}. While these models successfully capture protein spreading and tissue atrophy, how such structural degeneration translates into functional network degradation remains an open question. 

Recent studies suggest that the spatial distribution of tau pathology is closely associated with functional brain networks \citep{ossenkoppele_tau_2019,franzmeier_functional_2019,franzmeier_functional_2020, franzmeier_tau_2022}. Specifically, highly interconnected brain regions tend to exhibit similar patterns of tau accumulation, supporting the notion that tau pathology preferentially propagates along functional networks. As tau aggregation is known to induce structural damage—including cortical atrophy and white matter tract disruption—these pathological changes ultimately contribute to cognitive impairment \citep{bouzigues_structural_2025}. Alterations in functional connectivity are therefore plausibly linked to the disruption of underlying structural pathways. However, despite growing evidence connecting tau pathology to functional network organization, a clear qualitative or quantitative relationship between functional connectivity changes and tau-related downstream pathology—particularly tissue atrophy—remains insufficiently established. This gap is further complicated by methodological limitations in tau positron emission tomography (PET) imaging. In particular, the commonly used AV1451 tracer has been reported to exhibit off-target binding in regions such as the basal ganglia, hippocampus, and choroid plexus, potentially confounding the accurate quantification of regional tau burden \citep{marquie_pathological_2017, lemoine_tau_2018} . In contrast, regional brain volume changes, quantified as atrophy rates, can be robustly and non-invasively assessed in vivo using structural magnetic resonance imaging \citep{fox_imaging_2001, ridha_tracking_2006}. Functional connectivity alterations, in turn, can be independently measured using functional MRI. Integrating longitudinal atrophy metrics with functional connectivity analysis therefore provides a more reliable and biologically grounded framework for investigating how structural degeneration translates into functional network degradation. This perspective motivates the present study, which focuses on elucidating the relationship between atrophy rate and functional connectivity changes in Alzheimer’s disease.

Accordingly, the present study aims to establish a multiphysics modeling framework to reproduce tau protein propagation and brain atrophy patterns, and to ultimately elucidate the relationship between atrophy rate and functional connectivity alterations in Alzheimer’s disease. The model employs an anisotropic reaction–diffusion formulation to predict the spatiotemporal evolution of tau pathology, which is coupled with a biomechanical finite-deformation model to capture concentration-dependent tissue atrophy. Model performance is quantitatively evaluated by comparing simulated atrophy patterns with experimentally observed atrophy derived from imaging data. In parallel, longitudinal structural MRI and functional MRI data are analyzed, revealing a linear relationship between regional atrophy rates and functional connectivity changes. Based on this observation, regional atrophy rates extracted from the computational model are used to construct a structural network degradation matrix at the level of selected regions of interest. By integrating this degradation matrix into a neural oscillation framework, the model further predicts the corresponding trends in functional connectivity alterations, thereby providing a mechanistic link between protein-induced tissue degeneration and large-scale functional network disruption.
\fi

\section{Materials and Methods}
\ifshowMaterialsMethods
    \begin{figure}[H]
    \centering
    \includegraphics[width=0.9\linewidth]{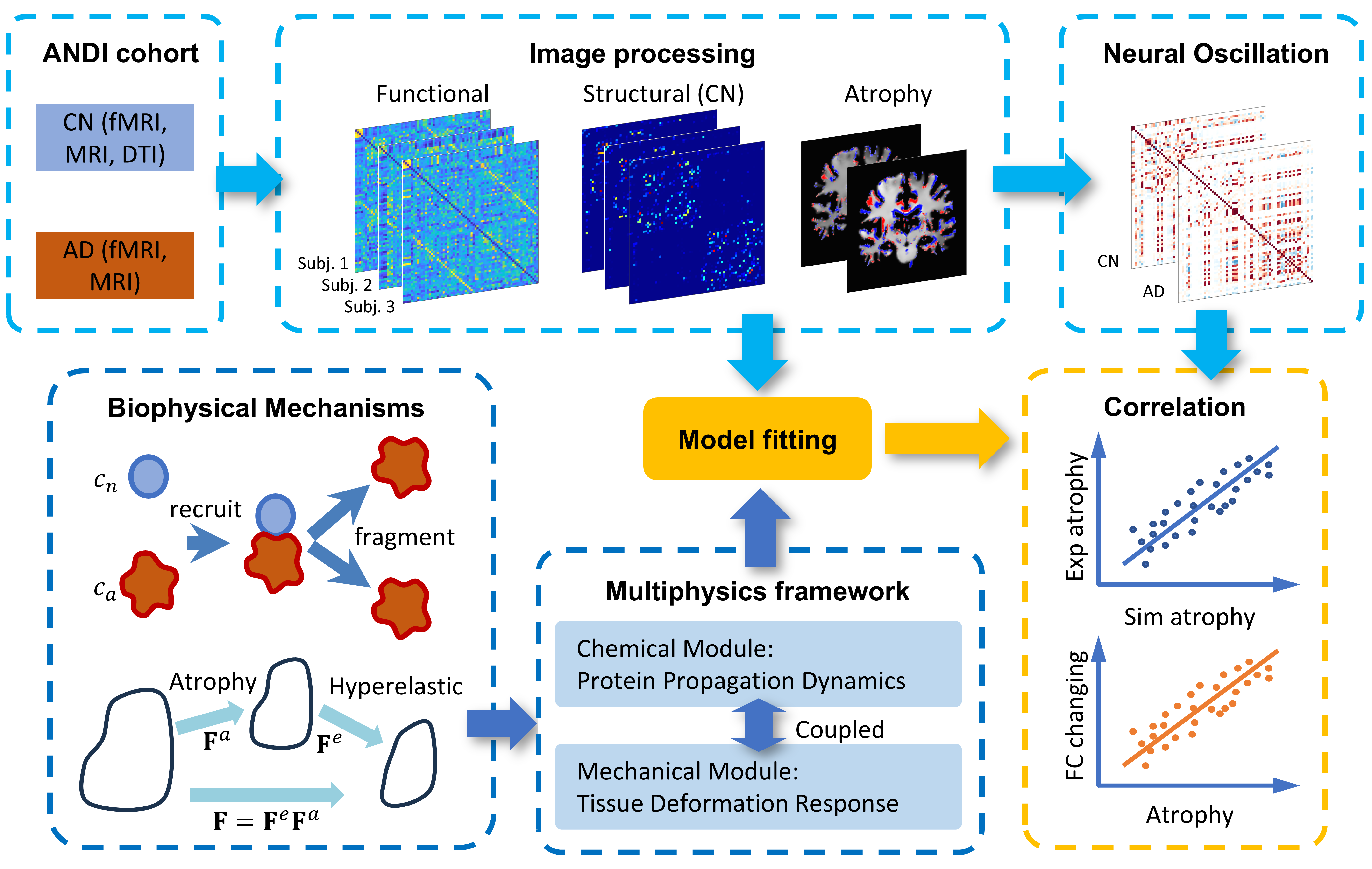}
    \caption{\textbf{An image-driven multiphysics workflow links tau propagation, atrophy, and functional network disruption.} Multimodal neuroimaging data from cognitively normal (CN) and Alzheimer’s disease (AD) subjects are processed to extract functional connectivity, structural information, and regional atrophy patterns. These image-derived measures inform a biophysically motivated multiphysics framework that couples tau protein propagation dynamics with hyperelastic tissue deformation to model whole-brain atrophy. In parallel, neural-oscillation dynamics are simulated and modulated by atrophy-induced degradation. Model parameters are calibrated through data-driven fitting, and the resulting simulations are quantitatively compared with imaging-derived measures to assess correlations between simulated and experimental atrophy as well as functional connectivity changes.}
    \label{fig: workflow}
\end{figure}
\subsection*{ADNI Cohort and Longitudinal Multimodal MRIs}
This study uses longitudinal magnetic resonance imaging (MRI) data obtained from the Alzheimer’s Disease Neuroimaging Initiative (ADNI) database. A total of 84 subjects were included, comprising 34 individuals diagnosed with Alzheimer’s disease (AD) and 50 cognitively normal (CN) controls. All selected subjects had multiple longitudinal structural MRI scans acquired over time, together with a final available resting-state functional MRI (fMRI) scan, enabling the joint analysis of structural atrophy and functional connectivity.

Diffusion MRI (dMRI) data were available for a subset of the cohort, consisting of 26 CN participants, and were used to derive representative white matter fiber orientations for anatomically informed modeling of anisotropic tau diffusion. The inclusion of dMRI data from cognitively normal subjects ensures that the embedded fiber architecture reflects baseline structural connectivity rather than disease-altered microstructure.

The average time interval between consecutive longitudinal MRI scans was approximately 1.0 year. For the AD group, follow-up durations ranged from 1.0 to 2.0 years, whereas the CN group exhibited a broader longitudinal span from 1.0 to 8.5 years. Although long-term longitudinal imaging remains limited in typical clinical cohorts, these data provide short-horizon constraints for estimating near-term regional atrophy rates and calibrating the model. The computational framework then leverages these observations to bridge temporal gaps and project disease dynamics beyond the available imaging window. All imaging data were preprocessed following standardized ADNI pipelines to ensure consistency across subjects, imaging sites, and acquisition protocols.

\subsection*{ROI-level Multimodal Image Data Analysis and Connectome Construction}
For each subject, longitudinal MRI, fMRI, and dMRI data were analyzed following the workflow summarized in \autoref{fig: workflow}. All imaging modalities were first co-registered to the MNI152 standard space using SPM \citep{ashburner2014spm12} and its structural processing toolbox CAT12 \citep{gaser_cat_2024}, employing fourth-degree spline interpolation to preserve anatomical accuracy during spatial normalization.

Subsequently, cortical surface reconstruction and volumetric segmentation were performed using FreeSurfer \citep{desikan_automated_2006, fischl_freesurfer_2012} on normalized MNI152 MRI images. This procedure partitions the brain into 68 cortical and 45 subcortical regions of interest (ROIs), enabling the extraction of voxel-wise and region-wise volumetric measures from each longitudinal MRI scan. Regional brain volumes were then tracked across time to compute subject-specific atrophy rates for each ROI.

For connectivity analysis, diffusion and functional data were further processed using DSI Studio \citep{yeh_dsi_2025}. dMRI data were used to reconstruct white matter fiber pathways and generate anatomically informed structural connectivity, while resting-state fMRI time series were employed to compute functional connectivity matrices based on ROI-averaged signals. The resulting connectivity matrices form the basis for subsequent network-level analysis and are consistent with established procedures in connectome-based studies \cite{schafer_network_2020}.

Together, this multimodal image processing pipeline enables the integrated quantification of regional atrophy, structural organization, and functional connectivity changes, providing essential inputs for validating the proposed multiphysics model and its predictions.

\subsection*{Multiphysics Model Linking Anisotropic Tau Propagation to Brain Atrophy}
A mechanistic description of Alzheimer’s disease progression requires explicit coupling between molecular pathology and macroscopic tissue degeneration. While reaction--diffusion models have proven effective in capturing the spatiotemporal spread of pathological proteins \citep{raj_network_2012,weickenmeier_multiphysics_2018}, they alone cannot account for the structural deformation induced by neurodegeneration. While multiphysics models coupling protein transport and tissue mechanics have been proposed, their ability to quantitatively translate tau-driven atrophy into functional network degradation remains limited. To address this gap, we adopt a unified multiphysics formulation that directly links tau propagation to mechanically consistent atrophy dynamics.

Neurodegeneration in Alzheimer’s disease unfolds as a coupled biochemical–mechanical process: misfolded tau spreads along brain pathways while simultaneously reshaping tissue geometry through progressive atrophy. Capturing this feedback is essential because structural deformation can in turn alter transport distances and local vulnerability, amplifying regional patterns of degeneration. In addition, tau pathology exhibits a strong spatial structure, preferentially advancing along white-matter pathways rather than spreading isotropically through the parenchyma. This directional bias motivates an anisotropic transport model informed by fiber architecture, coupled to a biomechanical description of tissue loss to capture how local tau burden drives progressive deformation. We therefore adopt a unified multiphysics chemomechanical formulation that integrates a tract-informed reaction–diffusion model for tau propagation with large-deformation brain mechanics, allowing direct translation from molecular pathology to tissue-level atrophy trajectories.

We model the spatiotemporal propagation of tau protein using a reaction--diffusion formulation that captures both spatial transport and local biochemical kinetics \citep{fisher1937wave,kolmogorov1937study}. Specifically, the evolution of misfolded tau protein concentration $c(\mathbf{x},t)$ is governed by a nonlinear reaction--diffusion equation,
\begin{equation}
    \dot{c} = \nabla \cdot (\mathbf{D} \nabla c) + \alpha c(1 - c),
    \label{eq:reaction-diff}
\end{equation}
where $\mathbf{D}$ denotes the diffusion tensor and $\alpha$ represents the local production rate of misfolded protein. The production term $\alpha c(1 - c)$ accounts for the combined effects of tau production, clearance, and pathological conversion, leading to saturation at high concentrations and preventing unbounded growth of protein accumulation.
To incorporate direction-dependent transport along neural pathways, the diffusion tensor is defined as 
\begin{equation}
    \mathbf{D}= D_p \mathbf{I} + D_a \mathbf{n}_0 \otimes \mathbf{n}_0,
\end{equation}
where $D_p$  and $D_a$ denote the passive (isotropic) and active (anisotropic) diffusion coefficients, respectively, and $\mathbf{n}_0$ represents the local fiber orientation. This formulation enables preferential tau propagation along white matter tracts while retaining isotropic diffusion in the surrounding tissue.

To model brain atrophy induced by tau pathology, tissue deformation is described through a multiplicative decomposition of the deformation gradient,
\begin{equation}
    \mathbf{F} = \mathbf{F}^{e}\mathbf{F}^{a} \quad \mathrm{and} \quad J=\det(\mathbf{F})=J^{e}J^{a},
\end{equation}
where $\mathbf{F}^e$ represents the elastic deformation and $\mathbf{F}^a$ encodes the inelastic volumetric change associated with tissue atrophy. Accordingly, $J^e = \det(\mathbf{F}^e)$ denotes the elastic volume change, while $J^a = \det(\mathbf{F}^a)$ quantifies the irreversible volume loss due to neurodegeneration.
Consistent with neuroanatomical observations, gray matter atrophy is assumed to be isotropic and is modeled as
\begin{equation}
    \mathbf{F}^{a} =\sqrt[3]{\theta } \mathbf{I} \quad \mathrm{and} \quad  \mathbf{F}^{e} =\mathbf{F}/\sqrt[3]{\theta},
\end{equation}
whereas white matter atrophy is assumed to be anisotropic, reflecting preferential thinning of fiber tracts orthogonal to the local fiber direction $\mathrm{n}_0$,
\begin{equation}
    \mathbf{F}^{a}= \sqrt{\theta}\mathbf{I}+(1-\sqrt{\theta})\mathrm{n}_{0}\otimes \mathrm{n}_{0} \quad \mathrm{(symmetrical)}.
\end{equation}
Here $\theta$ is a scalar atrophy parameter driven by the local tau protein concentration and evolving over time.

The model introduces a constitutive description of brain atrophy in which tissue undergoes a baseline, natural atrophy at a constant rate $G_{0}$. The accumulation of misfolded tau protein is assumed to accelerate this natural atrophy through a concentration-dependent amplification factor $\gamma(c)$. For simplicity, atrophy acceleration is activated once the protein concentration exceeds a critical threshold $c^{crit}$,
    \begin{align}
        \dot{\theta}_{c}=[1+\gamma(c)]G_{0}=
        \begin{cases}
        G_{0}, \quad &\mathrm{if} \quad c< c^{crit}
        \\
        G_{0}+G_{c}, \quad &\mathrm{if} \quad c\ge c^{crit}
        \end{cases}
        \quad \mathrm{where} \quad 
        \gamma(c)=\frac{G_{c}}{G_{0}}\mathcal{H} (c-c^{crit}),
    \end{align}
where $G_{c}$ denotes the tau-induced atrophy rate and $\mathcal{H}(\cdot)$ is the Heaviside function.
To characterize the mechanical response of brain tissue, a neo-Hookean free energy density is adopted. The atrophy-weighted elastic stored energy $\psi_{0}$ is defined in terms of the elastic
\begin{equation}
    \psi_{0}=J^{a}\psi \quad \mathrm{with} \quad
    \psi=\frac{1}{2}\mu[\mathbf{F}^{e}:\mathbf{F}^{e}-3-2\ln(J^{e})] + \frac{1}{2}\lambda\ln^{2}(J^{e}),
\end{equation}
where $\mu$ and $\lambda$ are the Lamé parameters. Here, we assign distinct Lamé parameters to gray and white matter to reflect their different mechanical stiffnesses. 
Specifically, we set $\mu = 2.07~\mathrm{kPa}$ and $\lambda = 101.43~\mathrm{kPa}$ for gray matter, and $\mu = 1.15~\mathrm{kPa}$ and $\lambda = 56.35~\mathrm{kPa}$ for white matter \citep{budday_mechanical_2017}. Following standard thermodynamic arguments, the corresponding Cauchy stress tensor $\mathbf{\sigma}$ is obtained as
\begin{equation}
    \boldsymbol{\mathrm{\sigma}}
    =\frac{1}{J}\frac{d\psi_0}{d\mathbf{F}}\mathbf{F}^{T}
    =\frac{1}{J^{e}} [ \mu \mathbf{F}^{e}(\mathbf{F}^{e})^{T} + \left(\lambda\ln(J^{e}) - \mu\right)\mathbf{I}].
\end{equation}
Together, these formulations establish a mechanistic coupling between tau protein propagation and tissue-level deformation, enabling the simultaneous simulation of biochemical transport, concentration-dependent atrophy, and large-deformation brain mechanics within a unified multiphysics framework.

\subsection*{Computational Whole-brain Discretization, Tract-informed Anisotropy, and Boundary Setup}
Reliable prediction of disease spread and deformation depends on a computational domain that preserves key anatomical boundaries and directional tissue organization. Whole-brain discretization enables the coupled transport–mechanics equations to be solved on realistic geometry, ensuring that both tau propagation pathways and regional tissue loss emerge from the model rather than being imposed. We therefore construct an anatomically resolved finite-element representation that distinguishes gray and white matter and supports tract-informed anisotropic transport.

\autoref{fig:mesh}.a illustrates the finite element discretization of the brain geometry, where dark gray and light gray regions correspond to gray matter and white matter, respectively. A hexahedral finite-element mesh was designed to  balance computational tractability with numerical fidelity. The mesh resolution is sufficiently refined to represent the smooth anatomical transition between gray and white matter while remaining efficient for the coupled transport–mechanics simulations.

To account for the preferential transport of tau protein along neural pathways, a tractography-guided fiber architecture derived from data of the Human Connectome Project Aging cohort, obtained via the DSI Studio data hub, is embedded into the finite element mesh \citep{yeh_dsi_2025}. The resulting fiber orientation field is mapped onto each integration point and used to define an orthotropic diffusion tensor, thereby enabling anisotropic active diffusion aligned with underlying white matter microstructure. This approach ensures that the modeled tau propagation is anatomically informed and reflects the directional characteristics of neural connectivity. Furthermore, \autoref{fig:mesh}.b illustrates the prescribed initial condition, in which a comparatively higher tau concentration is assigned within the entorhinal cortex to serve as a localized initiation region for the subsequent propagation process, consistent with established hypotheses of tau pathology onset. Specifically, we initialize a normalized tau concentration field $c(\textbf{x},0) \in [0,1]$  by prescribing an elevated value $c_{EC}^0$ inside the entorhinal cortex and a low baseline value $c_b^0$ elsewhere. In our simulations, we set $c_{EC}^0=0.8$ and $c_b^0=0$. In addition, the distal end of the brainstem is fixed to impose displacement boundary conditions and ensure mechanical well-posedness of the deformation problem.
\begin{figure}[H]
    \centering
    \includegraphics[width=0.9\linewidth]{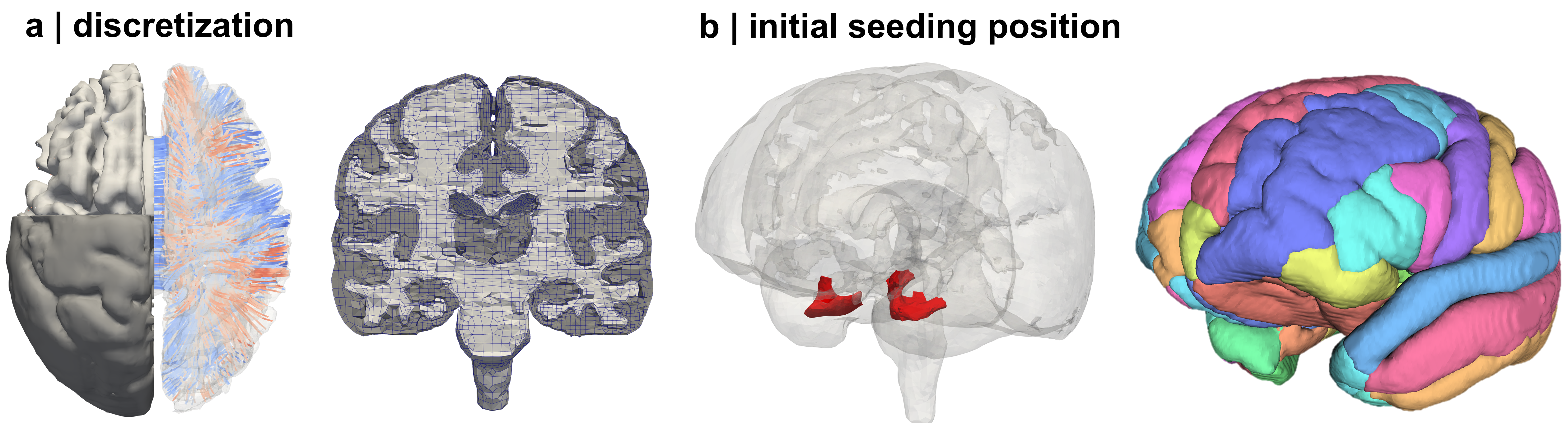}
    \caption{\textbf{An anatomically resolved finite-element mesh and entorhinal seeding define the initial conditions for tau propagation.} \textbf{a}, A hexahedral mesh distinguishes white matter (light gray) and gray matter (dark gray), and embedded fiber trajectories provide tract-informed orientations for anisotropic transport. \textbf{b}, Initial seeding regions in the entorhinal cortex that define the spatiotemporal onset of tau pathology. A total of 68 regions of interest—including 64 cortical regions along with the hippocampus and amygdala—are used to trace the subsequent progression.}
    \label{fig:mesh}
\end{figure}

\subsection*{Computational Implementation of Coupled Tau Transport and Deformation}
A coupled transport–mechanics model must be implemented in a way that preserves both numerical stability and physical interpretability, because small inconsistencies between the transport and deformation solvers can accumulate into large errors in predicted atrophy trajectories. We therefore adopt a single finite-element environment that enforces a consistent discretization, time integration, and parameter handling across biochemical transport and large-deformation mechanics. To achieve this efficiently, we leverage a heat–mass transport analogy that allows tau propagation to be solved using well-established thermal diffusion infrastructure while retaining full coupling to the mechanical response \citep{rayfield_individualized_2025}.

The proposed multiphysics framework is implemented within the commercial finite element software ABAQUS 2024 (Dassault Systems, Paris, France) using the user subroutines UMAT, UMATHT, and HETVAL. Due to the close mathematical analogy between tau protein transport and heat conduction, the UMATHT subroutine is employed to model the spatiotemporal evolution of tau protein concentration by exploiting the temperature degree of freedom. Specifically, both passive diffusion and orthotropic anisotropic active diffusion of tau protein are implemented through the conductivity tensor, enabling direction-dependent transport along preferred fiber orientations. The proliferation and clearance of tau protein are incorporated as volumetric source terms via the HETVAL subroutine, allowing flexible representation of local biochemical reactions. The resulting tau protein concentration field is treated as a global state variable and passed to the UMAT subroutine, where it governs the hyperelastic constitutive response of brain tissue, enabling concentration-dependent deformation within the mechanical module.
\subsection*{Neural Synchronization Model Linking Structural Atrophy to Functional Network Decline}
To quantify how structural atrophy translates into altered functional coupling, we require a network model that links changes in anatomical connectivity to emergent synchronization dynamics. Phase-oscillator frameworks provide a parsimonious yet mechanistically interpretable bridge between structure and function, and have been widely used to study coherence and desynchronization in large-scale brain networks. We therefore adopt a Kuramoto-type formulation and explicitly embed atrophy-dependent degradation into the coupling architecture to model progressive functional connectivity (FC) disruption.

In this framework, each ROI is modeled as a phase oscillator with its own intrinsic frequency and coupling strengths with other ROIs. According to the following evolution equation, the phase $\phi_i(t)$ of the $i$-th oscillator in a network of $N$ coupled oscillators evolves toward a coherent state through mutual interactions:
\begin{equation}
    \frac{d \phi_i}{dt} = \omega_i + g\sum_{j=1}^{N} K_{ji} \sin(\phi_i - \phi_j),
\end{equation}
where $\omega_i$ denotes the natural frequency of the $i$-th oscillator, drawn from a prescribed distribution, and $K_{ji}$ represents the coupling strength between the $j$-th and $i$-th oscillators, derived from the structural connectivity matrix. The effects of all connectivity
weights are scaled by a global coupling strength constant, $g$.

To obtain the simulated FC matrix, the stabilized structural connectivity matrix and the empirical functional connectivity matrix are jointly incorporated into an iterative optimization procedure. Specifically, the natural frequencies of the oscillators are adjusted based on network control energy. The initial natural frequencies are set to approximately 10~Hz, corresponding to neural oscillations during relaxed wakefulness, and are then updated to reduce the mismatch between simulated and empirical FC.

The brain network is further modeled as a controllable dynamical system governed by \citep{rayfield_individualized_2025}
\begin{equation}
    \frac{d\boldsymbol{x}(t)}{dt} = \mathbf{K}\boldsymbol{x}(t) + \mathbf{B}\boldsymbol{u}(t),
    \label{eq:neuroal_osci}
\end{equation}
where $\boldsymbol{x}(t)$ denotes the state vector representing functional connectivity, $\mathbf{B}$ is the control input matrix, and $\boldsymbol{u}(t)$ is an external control input that drives the system toward a target state.

To account for atrophy-induced disruption of structural connections-a process supported by evidence that gray matter atrophy often precedes and drives white matter disconnection \citep{fletcher_early_2014,lee_tract-based_2015}-we introduce a degradation matrix $\mathbf{M}$ constructed from the averaged atrophy degree $\boldsymbol{\theta}$ within each ROI:
\begin{equation}
    M_{ij} = \theta_i \theta_j,
    \label{degra_matrix}
\end{equation}
where $M_{ij}$ represents the degradation factor associated with the connection between the $i$-th and $j$-th ROIs, and $\theta_i$ and $\theta_j$ denote the averaged atrophy extent of the corresponding ROIs. This formulation assumes that the atrophy of both ROIs jointly contributes to the degradation of the inter-regional connection, aligning with neuroimaging findings that white matter tract integrity is closely correlated with the volume and neurodegeneration of anatomically connected gray matter regions \citep{lee_tract-based_2015,douaud_brain_2013}. The degradation matrix $\mathbf{M}$ is subsequently incorporated into the network dynamics to model functional connectivity under neurodegenerative effects. Accordingly, Eq.\ref{eq:neuroal_osci} can be rewritten as
\begin{equation}
    \frac{d\boldsymbol{x}(t)}{dt} = \mathbf{M}\mathbf{K}\boldsymbol{x}(t) + \mathbf{B}\boldsymbol{u}(t).
    \label{eq:neuroal_osci_degra}
\end{equation}

To ensure biological plausibility in the adjustment of network dynamics, the transition from an initial state $\boldsymbol{x}_0$ (simulated FC) to a target state $\boldsymbol{x}_f$ (empirical FC) is assumed to follow a minimum-energy principle. The control energy is defined as
\begin{equation}
    CE = \int_{0}^{T} \|\boldsymbol{u}(t)\|^2 \, dt.
\end{equation}
Accordingly, the natural frequency of the $i$-th ROI is updated based on both the control energy and the discrepancy between simulated ($sFC_i$) and empirical ($FC_i$) functional connectivity:
\begin{equation}
d\omega_i = CE \cdot \left(\sum_{j=1}^{n} sFC_{ij} - \sum_{j=1}^{n} FC_{ij}\right).
\end{equation}
where $sFC$ is obtained from the Kuramoto model. The updated natural frequency at the $(n+1)$-th iteration is given by
\begin{equation}
    \omega_i^{n+1} =
    \begin{cases}
        |\omega_i^{n} + d\omega_i|, 
        & \text{if } \omega_i^{n} - \mathrm{mean}(\boldsymbol{\omega}) \ge 0, \\[6pt]
        |\omega_i^{n} - d\omega_i|, 
        & \text{otherwise}.
    \end{cases}
\end{equation}

\fi
\section{Results}
\ifshowresults
    \subsection*{Edge-wise Statistics Identify Widespread Functional Connectivity Degradation}
Functional connectivity (FC) disruption is a hallmark of Alzheimer’s disease and provides a systems-level readout of how neurodegeneration reshapes communication across distributed brain networks. Establishing the magnitude and topography of these changes is essential for two reasons: it motivates the need for mechanistic structure–function coupling, and it provides an empirical target against which model-predicted network degradation can be evaluated. We therefore first characterize group-level FC differences between cognitively normal (CN) controls and individuals with Alzheimer’s disease (AD).

Derived from fMRI of two groups of subjects, the group-level averaged FC matrix of the CN and AD groups is obtained as shown in \autoref{fig:func_degra}.a. Due to a lack of longitudinal fMRI scans, subsequent analyses focused on cross-sectional differences between group means. \autoref{fig:func_degra}.b exhibits the differences between them, which demonstrate pronounced degradation in most regions of interest, except possibly compensatory in a few ROI \citep{han_functional_2009, penalba-sanchez_increased_2022}. Apart from ROI-wise mean changes, we assessed the significant difference between the two FC matrices at the element-wise level. For each target ROI, we extracted its connectivity strength with all other ROIs for every participant in the AD and CN groups. Group differences were evaluated using two-sample t-tests applied to the connectivity values $\mathrm{FC}_{AD}(i,j)$, $\mathrm{FC}_{CN}(i,j)$ across subjects, where $i$ denotes the target ROI and $j$ the guest ROI. ROI pairs with uncorrected $p < 0.05$ were considered to show significant group differences in connectivity, and the corresponding FC edge was identified according to these values, as shown in \autoref{fig:func_degra}.c (edges with color). Based on the significant FC changing matrix, ROI-wise FC connections are visualized in inter-hemisphere (\autoref{fig:func_degra}.d) and intra-hemisphere (\autoref{fig:func_degra}.e,f), respectively (the blue chord indicates a degraded connection, versus increasing). The connection between the inner of the left and right hemispheres both exhibit overall degradation, and different degradation patterns appear in the left and right intra-hemispheres.
\begin{figure}
    \centering
    \includegraphics[width=0.9\linewidth]{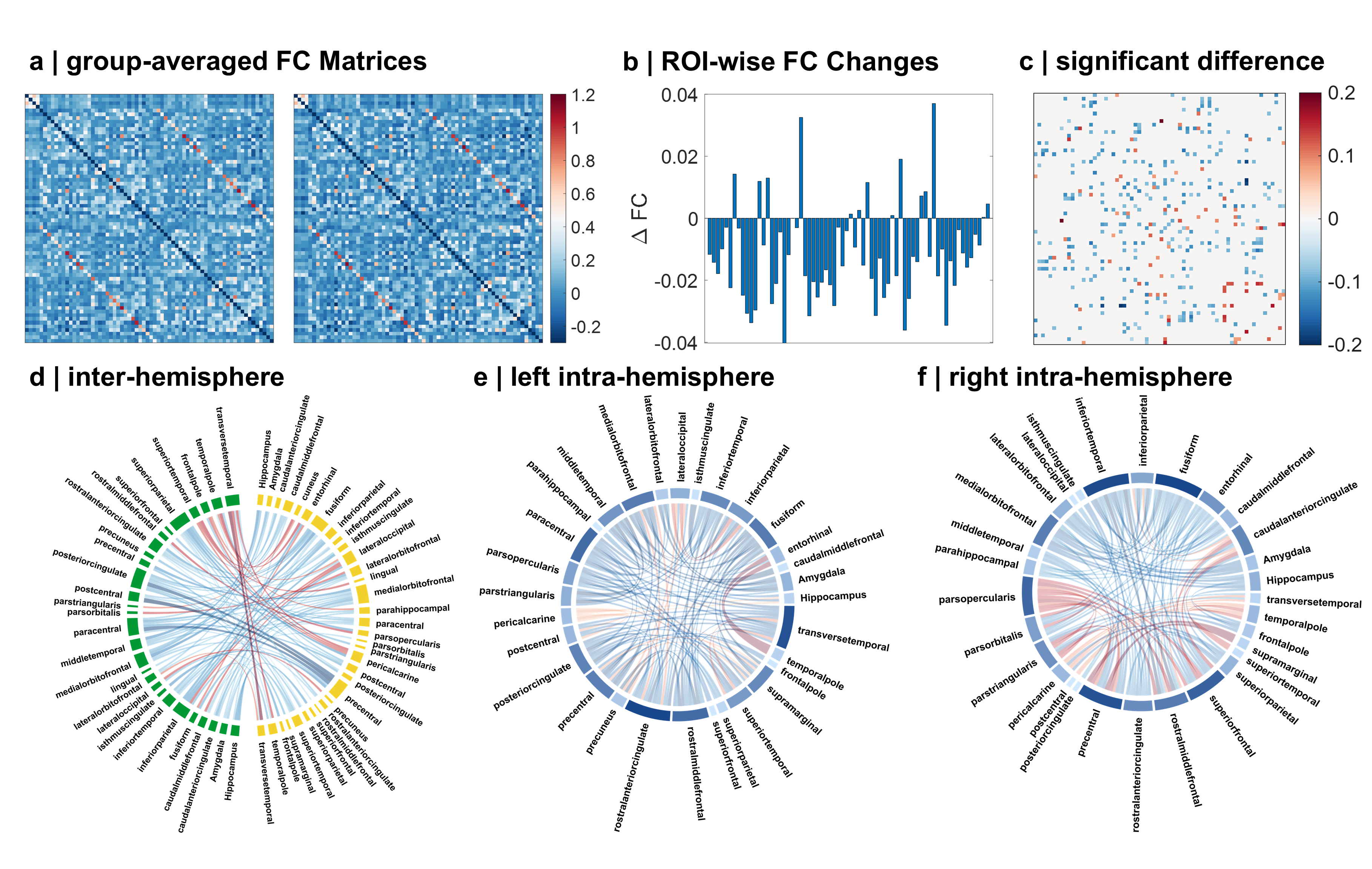}
    \caption{\textbf{Group-level functional connectivity differs systematically between cognitively normal controls and Alzheimer’s disease.} \textbf{a}, Group-averaged functional connectivity (FC) matrices are shown for cognitively normal (CN) and Alzheimer’s disease (AD) cohorts. \textbf{b}, ROI-wise changes in FC  indicate an overall reduction across most regions. \textbf{c}, ROIs exhibiting statistically significant between-group differences (two-sample t-test, uncorrected $p<0.05$) are highlighted in the difference matrix. \textbf{d}, Inter-hemispheric connections derived from the significant-difference matrix reveal altered cross-hemisphere coupling, where blue indicates decreased FC and red indicates increased FC in AD relative to CN. \textbf{e,f}, Left and right intra-hemispheric patterns further show hemisphere-specific topological reorganization.}
    \label{fig:func_degra}
\end{figure}
\subsection*{Structure–Function Coupling Links Accelerated Atrophy to Network-Level Degeneration}
Linking progressive tissue loss to network-level dysfunction is central to understanding why Alzheimer’s disease evolves from a regional pathology into a distributed systems disorder. Longitudinal atrophy provides a quantitative readout of structural degeneration, but its functional consequences are often heterogeneous across regions, reflecting both loss of anatomical support and potential compensatory reorganization. We therefore first characterize ROI-specific atrophy trajectories and then test whether regional atrophy rates systematically covary with functional connectivity (FC) changes.

\autoref{fig:atrophy_func_coupling} illustrates the longitudinal trajectories of brain volume loss and its association with variations in functional connectivity. We extracted the cortical, hippocampal, and amygdalar volumes based on the atlas defined in \autoref{fig:mesh}.b. To enable inter-subject comparison, all volumes are normalized to each subject's baseline measurement. In \autoref{fig:atrophy_func_coupling}.a, the longitudinal trajectories of individual subjects in the two groups are shown, with solid lines indicating the group-level linear regression fits. Each data point corresponds to a single MRI scan. The results reveal a clear divergence in volume decline between the CN and AD groups, with AD exhibiting accelerated decline. By performing linear regression within each ROI, we further derived the atrophy rates, which are presented in \autoref{fig:atrophy_func_coupling}.b. The pronounced group difference (***$p<0.001$) in atrophy rates indicates that tau pathology markedly accelerates regional brain volume loss. To further investigate the coupling between structural degeneration and functional alterations, we quantified the FC difference and atrophy-rate difference in each ROI. Specifically, for every ROI, we first computed the mean FC of the AD group and subtracted the corresponding mean FC of the CN group, yielding the FC difference ($\mathrm{FC}_{AD} - \mathrm{FC}_{CN}$). Likewise, the atrophy difference for each ROI was obtained by subtracting the CN-group atrophy rate from the AD-group atrophy rate. We then calculated the Pearson correlation between these two ROI-wise variables. The resulting significant association shows in \autoref{fig:atrophy_func_coupling}.c that regions exhibiting larger tau-induced increases in atrophy rate also present greater FC degradation. However, a subset of ROIs displays the opposite pattern: despite comparatively small atrophy rates, these regions show increased FC. Such FC enhancement is likely reflective of compensatory functional reorganization in response to early or mild structural decline, a phenomenon that has been reported in several neurodegenerative studies and is often interpreted as short-term network-level adaptation aimed at maintaining cognitive performance. 
\begin{figure}
    \centering
    \includegraphics[width=0.9\linewidth]{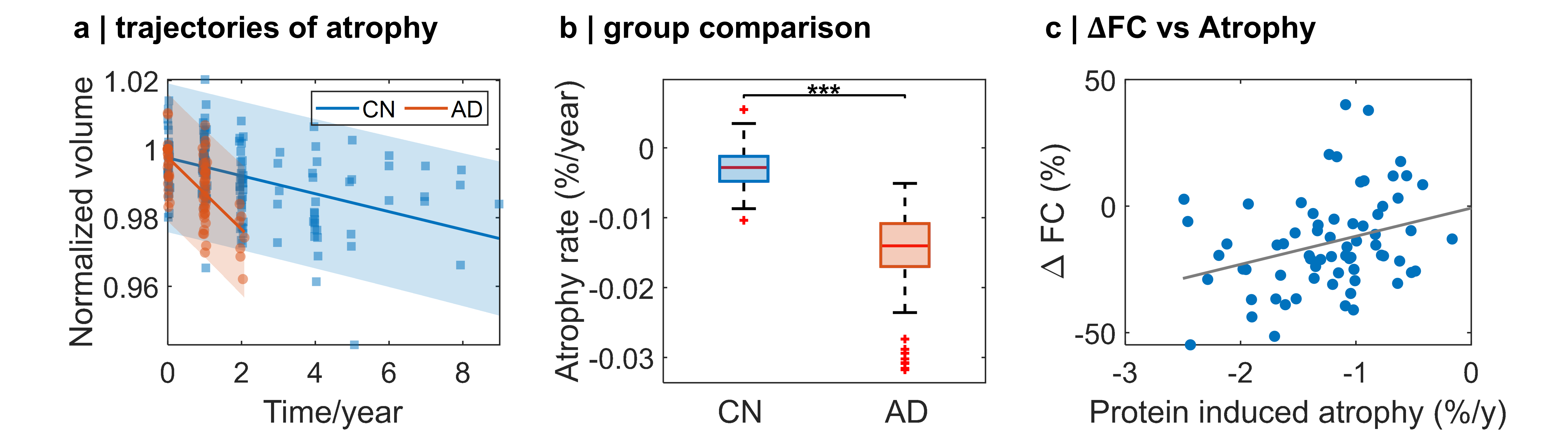}
    \caption{\textbf{Longitudinal atrophy rates are associated with cross-sectional functional connectivity differences.} \textbf{a}, Longitudinal trajectories of normalized brain volume for cognitively normal (CN) and Alzheimer’s disease (AD) groups, with regression lines and confidence intervals illustrating faster atrophy in the AD cohort. \textbf{b}, Atrophy rates for CN and AD participants, showing a significantly accelerated decline in the AD group (***$p < 0.001$). \textbf{c}, Across ROIs, protein-induced increases in atrophy rate are positively correlated with FC changes ($r=0.31$, $p=0.0103$), indicating that stronger structural degeneration is accompanied by larger functional disruption.}
    \label{fig:atrophy_func_coupling}
\end{figure}
\subsection*{Model Predictions Reproduce Hallmark Spatiotemporal Patterns of Tau Spread and Atrophy}
A key requirement for any mechanistic model of Alzheimer’s disease is that it reproduces the characteristic timing and topology of pathology observed in vivo. Capturing not only the overall burden of tau and tissue loss, but also their spatial progression, is essential for establishing biological plausibility and for enabling downstream predictions of network-level disruption. We therefore evaluate whether the proposed multiphysics framework can recapitulate canonical neurodegenerative signatures of tauopathies.

Based on the proposed multiphysics framework, the coupled spatiotemporal evolution of pathological tau protein and progressive brain atrophy is reproduced, as illustrated in \autoref{fig:avg_results}. In the left panel, the whole-brain–averaged volume decreases monotonically over time, whereas the mean protein concentration increases in a nonlinear, accelerating manner, indicating a progressively worsening pathological burden. The simulated temporal trajectories of whole-brain–averaged volume and protein concentration capture hallmark features of neurodegenerative progression, exhibiting trends that are qualitatively consistent with clinical and imaging observations of tauopathies. The coronal slices in the right panel of \autoref{fig:avg_results} provide a detailed view of the spatial evolution of tau pathology over time from Year 1, 4, 7, to Year 10. At early stages, elevated tau concentrations are primarily localized within the entorhinal region, reflecting known sites of initial pathological onset. As the disease progresses, tau accumulation expands into adjacent cortical regions and subsequently propagates deeper into the temporal lobe.

Notably, the spatial distribution of tau demonstrates pronounced anisotropy, with preferential propagation aligned with major fiber pathways. This behavior highlights the dominant role of axonal transport in governing pathological protein spread. The resulting fiber-aligned propagation patterns reinforce the importance of white-matter connectivity in shaping disease progression and validate the necessity of incorporating structural anisotropy within the modeling framework.
\begin{figure}
    \centering
    \includegraphics[width=0.9\linewidth]{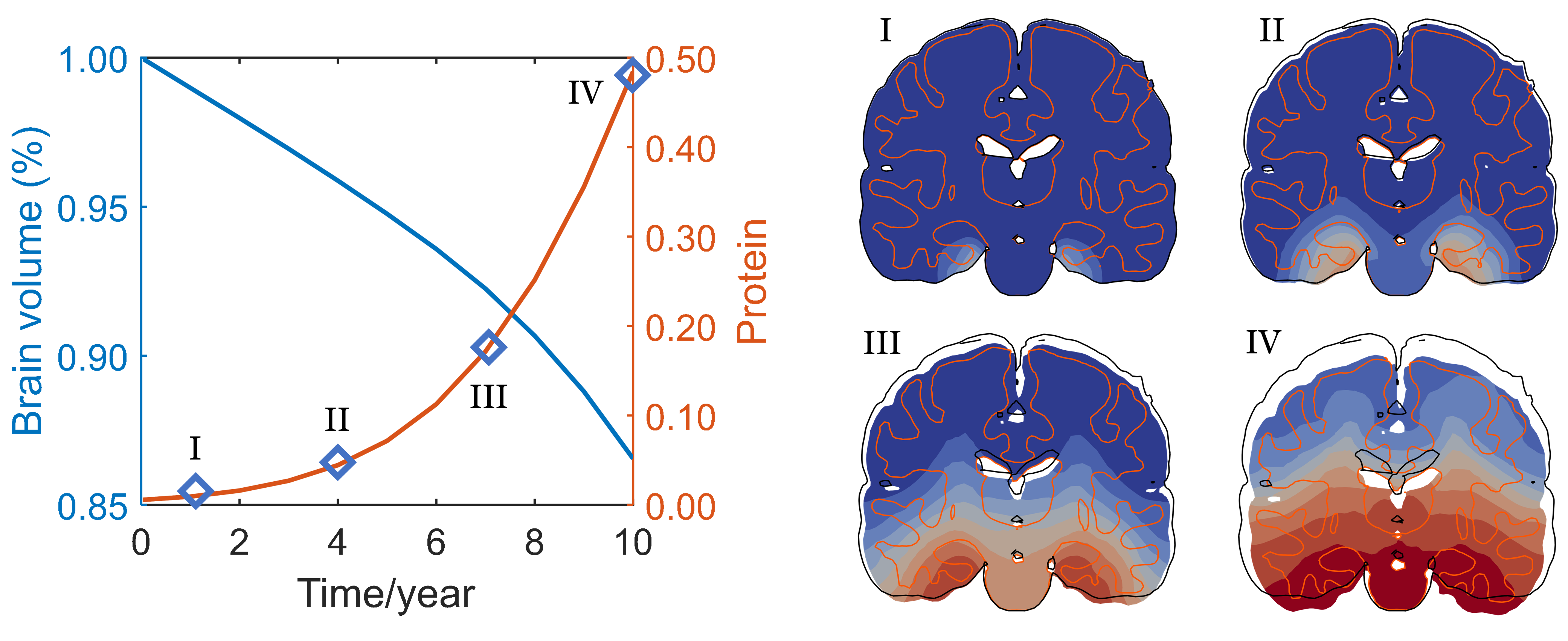}
    \caption{\textbf{Coupled tau transport and large-deformation mechanics reproduce whole-brain atrophy and evolving protein burden.} The left panel shows simulated temporal trajectories of normalized brain volume and total tau burden over a 10-year period. The right panels show representative coronal slices at 1, 4, 7, and 10 years, illustrating the spatial distribution of protein concentration. Overlaid contours depict morphological deformation relative to baseline anatomy, and the white--gray matter interface is delineated by a solid line.}
    \label{fig:avg_results}
\end{figure}
\subsection*{Predicted Long-Term Atrophy Trajectories Explain Downstream Functional Connectivity Decline}
Longitudinal neuroimaging typically spans only a few years, yet Alzheimer’s disease unfolds over a decade or more. Mechanistic models are therefore most valuable when they can extrapolate beyond the observation window while preserving known disease topology and providing quantitative, testable predictions. We used the multiphysics framework to forecast long-term regional atrophy and to determine whether these model-predicted structural changes are sufficient to explain downstream functional connectivity (FC) degradation.

Because available longitudinal MRI covered a limited duration, we simulated extended trajectories of regional volume loss. \autoref{fig:model_predict}.a presents the evolution of volume loss across all ROIs in the left and right hemispheres. While both hemispheres show broadly similar trajectories, subtle asymmetries emerge in several regions, reflecting underlying differences in fiber-tract architecture. The entorhinal cortex, hippocampus, and amygdala demonstrate the most pronounced atrophy, aligning well with the canonical sequence of degeneration described in Braak staging. \autoref{fig:model_predict}.b–d depicts the spatial distribution of white-matter volume loss at 4, 7, and 10 years. The solid black contours denote the baseline white-matter boundaries, highlighting the progressive inward shrinkage of tissue over time. The predicted atrophy pattern closely mirrors the spread of tau pathology: regions adjacent to the entorhinal cortex exhibit the earliest and most severe degeneration, and the white matter as a whole undergoes substantial shrinkage as the disease advances.
\begin{figure}
    \centering
    \includegraphics[width=0.9\linewidth]{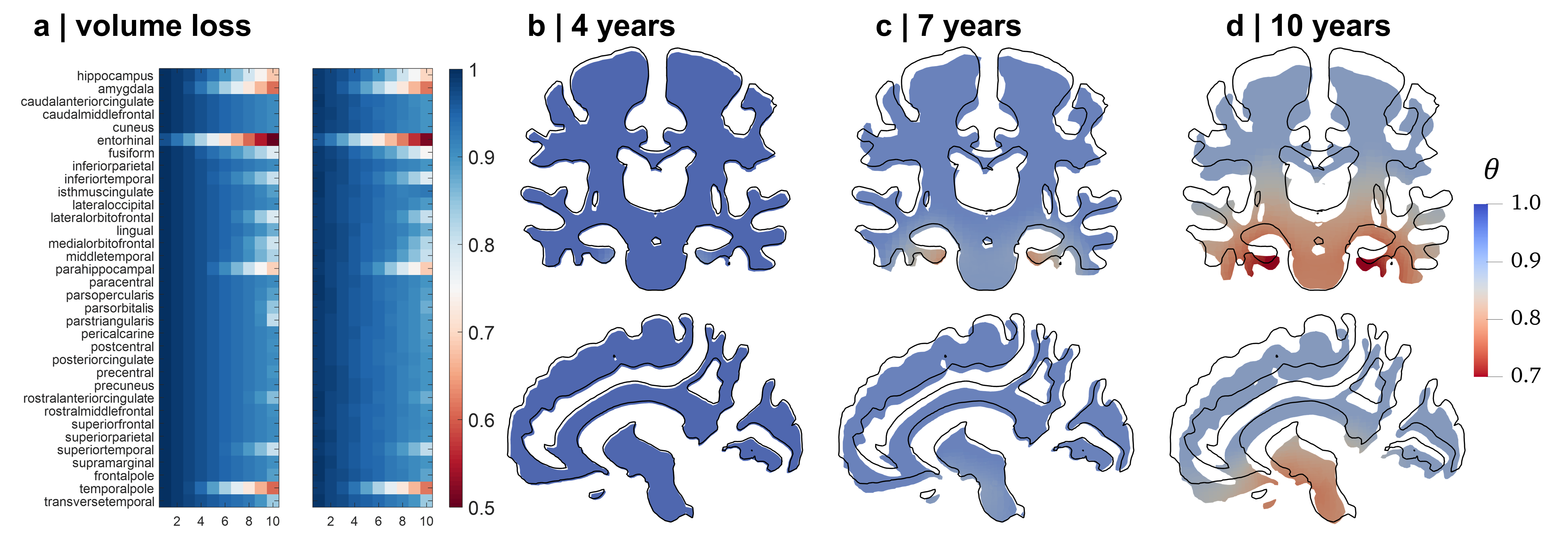}
    \caption{\textbf{The multiphysics model predicts ROI-resolved atrophy trajectories and spatially heterogeneous tissue loss.} \textbf{a}, Heatmaps showing the longitudinal trajectories of normalized regional volume across 68 brain regions over 10 years in the left and right hemispheres, respectively. Warmer colors indicate more severe volume loss. \textbf{b-d}, Model-predicted spatial patterns of tissue loss visualized on coronal (top) and sagittal (bottom) slices at three representative time points. The overlaid contours show progressive deformation of white-gray matter boundaries relative to baseline anatomy.}
    \label{fig:model_predict}
\end{figure}
We next benchmarked model predictions against imaging-derived measurements. As shown in \autoref{fig:atrophy_exp-sim}, the model effectively captures experimentally observed atrophy patterns and reproduces the empirical relationship between functional connectivity (FC) changes and protein-induced degeneration. \autoref{fig:atrophy_exp-sim}.a demonstrates that simulated atrophy rates across ROIs closely match the experimental AD measurements, yielding a strong correlation ($r=0.81, p=5.5\times10^{-17}$). The high consistency in regions exhibiting peak atrophy further indicates that the multiphysics framework can quantitatively reproduce the spatial distribution of degeneration. \autoref{fig:atrophy_exp-sim}.b shows a comparable positive association between simulated protein-induced atrophy and FC alterations. Notably, regions with relatively low atrophy rates exhibit detectable FC enhancement, consistent with the aforementioned compensatory phenomenon. To further examine model performance, \autoref{fig:atrophy_exp-sim}.c compares simulated and experimental trajectories within three representative regions—entorhinal cortex, hippocampus, and amygdala. Differences between the left and right hemispheres are also evaluated (solid line: left; dashed line: right), and the inset renderings illustrate final-time ROI volumes with baseline contours overlaid. To more clearly reveal the model’s predictive accuracy, comparisons are made starting from year 4 of the simulation, after the tau pathology has expanded beyond the immediate vicinity of the entorhinal cortex. Across all three ROIs, the model successfully reproduces both the magnitude and slope of atrophy observed in AD participants, as well as the hemispheric asymmetry in degeneration. In particular, the hippocampus exhibits distinct left–right differences that align with heterogeneous fiber architecture—differences that the model accurately predicts through its integration of structural connectivity.
\begin{figure}
    \centering
    \includegraphics[width=0.9\linewidth]{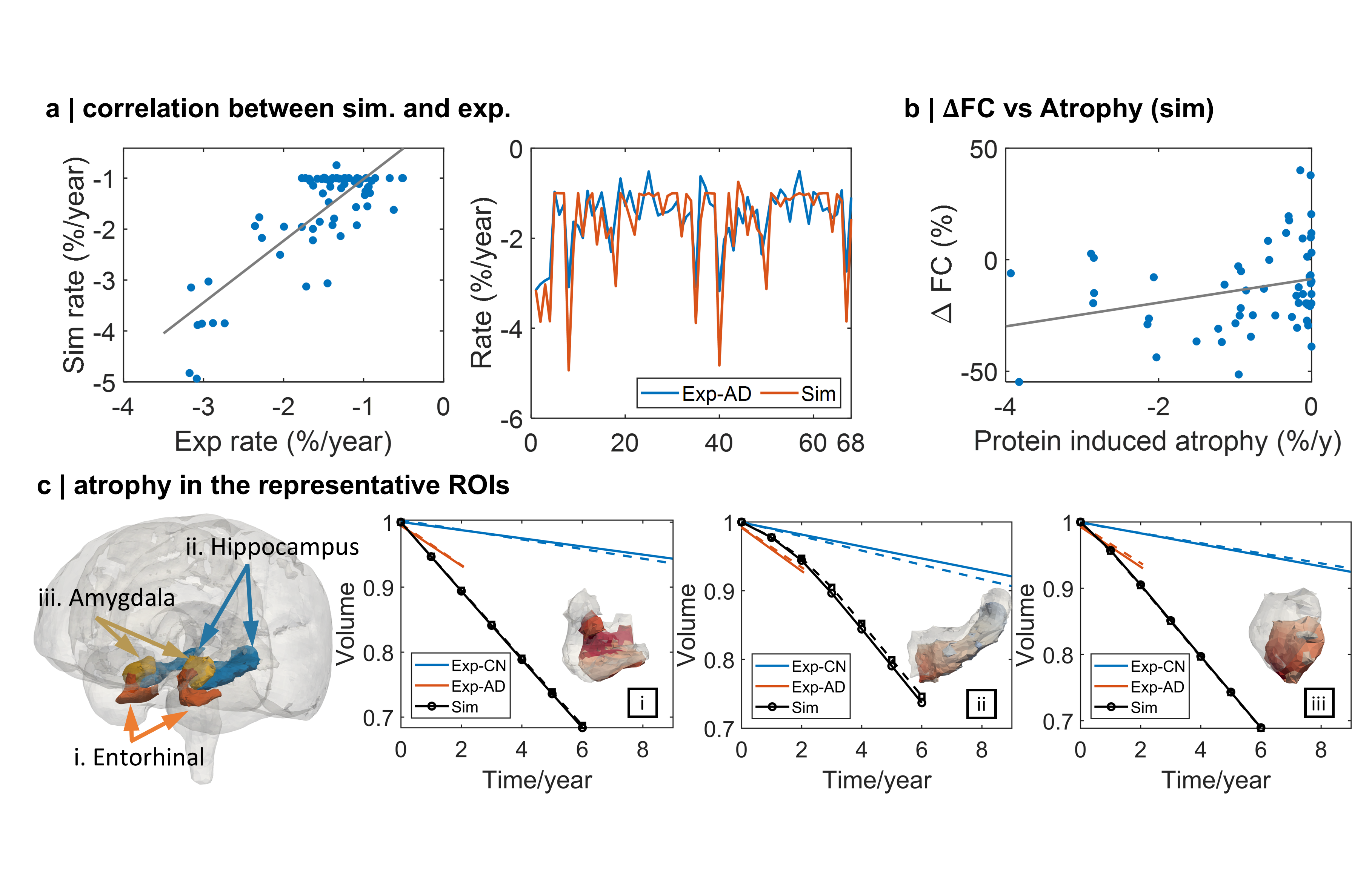}
    \caption{\textbf{Simulated atrophy rates match imaging-derived measurements and reproduce the empirical atrophy--FC association.} \textbf{a}, Simulated and experimental ROI-wise atrophy rates show strong agreement ($r=0.81$, $p=5.5\times10^{-17}$). And rate differences over all brain regions for experimental (Exp–AD) and simulated (Sim) data demonstrate the great capacity of the model to capture the atrophy patterns. \textbf{b}, Simulated protein-induced atrophy rates are positively associated with FC changes across ROIs ($r=0.28$, $p=0.04$). \textbf{c}, Representative trajectories in the entorhinal cortex, hippocampus, and amygdala show close agreement between simulated and experimental atrophy (solid line: left hemisphere; dashed line: right hemisphere).}
    \label{fig:atrophy_exp-sim}
\end{figure}
As shown in \autoref{fig:func_degra_exp-sim}, we simulate functional connectivity (FC) based on the structural connectivity (SC) matrix (\autoref{fig:func_degra_exp-sim}.a) constructed from HCP-A DTI data, and subsequently examine the relationship between simulated FC changes and model-predicted atrophy. To incorporate the impact of structural degeneration, we introduce a degradation matrix $\mathbf{M}$, derived from the regional atrophy rates. The FC of the CN group is directly simulated from the intact SC, in \autoref{fig:func_degra_exp-sim}.b, whereas the FC of the AD group is generated from the SC matrix modified by $\mathbf{M}$. \autoref{fig:func_degra_exp-sim}.c highlights the difference between the AD and CN FC matrices (AD - CN). A widespread reduction in FC strength is observed across most edges, consistent with the global degradation in AD. The ROI-wise FC changes in \autoref{fig:func_degra_exp-sim}.d further show an overall decline across regions. One clear outlier (marked in red) exhibits an apparent FC decrease in the simulation; however, the corresponding experimental FC change is positive. As shown in \autoref{fig:func_degra_exp-sim}.e, this region has an atrophy rate close to zero, suggesting that the anomalous FC estimate likely arises from the intrinsic properties of the neural oscillator model rather than from structural degeneration, and can thus be reasonably excluded from the correlation analysis. After removing the outlier, the resulting association between simulated FC change and protein-induced atrophy rate remains highly significant and closely matches the empirical trend shown in \autoref{fig:atrophy_func_coupling}. Importantly, the model also reproduces FC enhancement in regions with minimal atrophy, mirroring experimental observations. These findings demonstrate that the proposed multiphysics framework not only predicts atrophy trajectories with high fidelity but also accurately captures the downstream degradation of functional connectivity. The introduction of the degradation matrix $\mathbf{M}$ further provides mechanistic insight into how structural decline shapes functional disconnection across the brain's network architecture.
\begin{figure}
    \centering
    \includegraphics[width=0.9\linewidth]{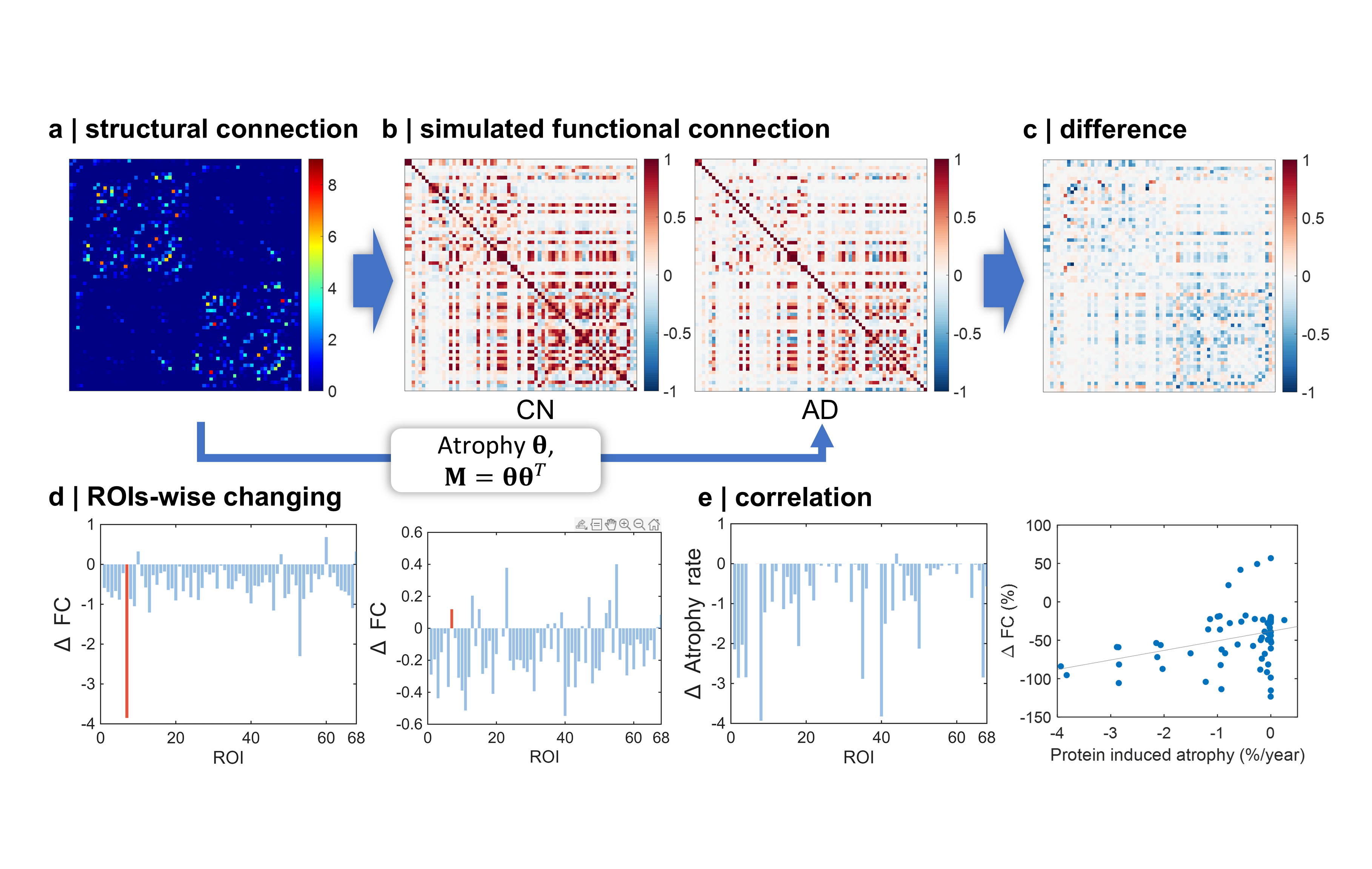}
    \caption{\textbf{Atrophy-informed degradation of structural coupling predicts functional connectivity disruption.} \textbf{a}, Group-averaged structural connectivity (SC) matrix used as the baseline coupling topology for all simulations. \textbf{b}, Simulated functional connectivity (FC) matrices for cognitively normal controls (CN) and Alzheimer’s disease (AD). An atrophy-derived degradation vector $\boldsymbol{\theta}$ is used to construct a rank-one degradation matrix $\mathbf{M}=\boldsymbol{\theta}\boldsymbol{\theta}^{\mathsf T}$, which modulates (down-weights) the effective SC coupling strengths in the neural-oscillation model to generate AD-like FC patterns. \textbf{c}, Difference matrix between simulated AD and CN FC shown only for ROI pairs with statistically significant differences; non-significant entries are masked in white. Red indicates increased FC, whereas blue indicates decreased FC in AD relative to CN. \textbf{d}, ROI-wise FC changes ($\Delta$FC) derived from the simulated CN and AD networks, with red-highlighted bars denoting outlier ROIs. \textbf{e}, Association between protein-induced regional atrophy rates and FC changes: ROI-wise atrophy rate estimates (left) and their correlation with $\Delta$FC across regions (right), demonstrating that stronger atrophy is linked to larger FC degradation ($r=0.34, p=0.0045$).}

    \label{fig:func_degra_exp-sim}
\end{figure}
\fi
\section{Discussion}
\
\ifshowDiscussion
    This study asked a focused question: whether Alzheimer’s disease (AD)-related functional connectivity (FC) disruption is systematically associated with tau-linked neurodegeneration, and whether such a relationship can be mechanistically instantiated in a computational model. Using a cross-sectional group contrast (healthy controls vs. AD), we observed significantly reduced FC in AD, consistent with extensive evidence of resting-state network disruption and loss of coordinated large-scale communication in the disease \citep{greicius_default-mode_2004,brier_loss_2012,badhwar_restingstate_2017,seeley_neurodegenerative_2009}. Beyond confirming group-level FC impairment, our key result is that the spatial pattern of FC degradation across ROIs is positively associated with tau-induced atrophy parameters derived from our multiphysics framework. This association motivated a parsimonious degeneration-matrix formulation, which qualitatively reproduced the AD-like FC disruption pattern when applied to the baseline structural scaffold, thereby providing a mechanistic bridge from molecular pathology and tissue loss to systems-level dysfunction.

\subsection*{Mechanistic Interpretation: from Tau Pathology to Network Disconnection}
A biologically grounded interpretation of our findings is that tau-driven tissue loss captures, in aggregate form, the progressive dismantling of synaptic and axonal substrates required for long-range coordination. Neuropathological staging and in vivo PET studies consistently show that tau pathology follows a stereotyped topographic expansion beginning in transentorhinal/entorhinal regions and advancing to limbic and association cortices \citep{braak_neuropathological_1991,ossenkoppele_tau_2016}. A large experimental literature further supports the idea that pathological tau can propagate in a ``prion-like'' manner along connected neuronal pathways, providing a natural mechanistic link between connectivity architecture and regional vulnerability \citep{clavaguera_transmission_2009,decalignon_propagation_2012,goedert_like_2017,jucker_propagation_2018}. Critically, synapse loss and disconnection are among the strongest proximate correlates of cognitive impairment in AD, reinforcing the view that macroscopic neurodegeneration ultimately manifests as impaired inter-regional communication \citep{dekosky_synapse_1990,terry_physical_1991}. At the mesoscale, diffusion MRI studies report anatomically congruent associations between white-matter microstructural abnormalities and gray-matter atrophy, consistent with coupled gray--white degeneration and tract-level compromise \citep{agosta_white_2011}. Taken together, these findings support our modeling choice: using ROI-level atrophy parameters as a mathematically tractable proxy for the cumulative loss of effective coupling capacity that underlies FC disruption.

\subsection*{Computational Novelty: Explicit SC--FC Coupling under Tau-linked Degeneration}
Prior connectome-based propagation models have largely focused on explaining or forecasting spatial patterns of pathology/atrophy, whereas here we explicitly model how tau-related atrophy parameters reshape FC, providing a mechanistic bridge from molecular pathology to systems-level dysfunction \citep{raj_network_2012,iturria-medina_epidemic_2014,franzmeier_functional_2020}. In particular, diffusion- and epidemic-inspired models demonstrate that network topology and anatomical connectivity can constrain where misfolded proteins accumulate and where neurodegeneration emerges \citep{raj_network_2012,iturria-medina_epidemic_2014}. Complementing these approaches, recent in vivo work shows that functional architecture is closely linked to tau accumulation and spatial tau organization, reinforcing the hypothesis that connectivity is not merely a passive background but a determinant of tau dynamics \citep{franzmeier_functional_2019,franzmeier_functional_2020}. Our contribution is to leverage these insights in the reverse direction: rather than stopping at reproducing tau/atrophy maps, we use multiphysics-derived atrophy parameters to build a low-dimensional, interpretable degeneration operator that modulates the structural connectome and thereby generates disease-state FC patterns.
From a dynamical-systems perspective, this step is well motivated. Whole-brain modeling has repeatedly shown that resting-state FC can be partially explained as an emergent property of neural dynamics constrained by the structural connectome \citep{honey_predicting_2009}. In particular, coupled-oscillator and related neural-mass models demonstrate how connectome-mediated coupling and local dynamics can reproduce salient FC patterns \citep{cabral_role_2011,kuramoto_chemical_1984,deco_identification_2014}. Within this modeling class, our degeneration matrix provides a minimal mechanism by which tau-linked tissue loss can be translated into a systematic penalty on inter-regional coupling. The fact that such a simple rule can qualitatively reproduce the observed AD-like FC disruption pattern suggests that a substantial component of cross-sectional FC differences may be attributable to degeneration-induced weakening of effective coupling.

\subsection*{Network Specificity and Clinical Relevance}
Our results indicate that FC alterations in AD are spatially structured rather than random. Medial temporal and limbic areas (e.g., entorhinal cortex, hippocampus/parahippocampus, temporal pole, and amygdala) are well-established early sites of tau pathology and neurodegeneration \citep{braak_neuropathological_1991,maass_entorhinal_2018,ossenkoppele_tau_2016}. Notably, FC change need not be uniformly negative within these tau-vulnerable regions; instead, our robust finding is a cross-ROI relationship in which ROIs with larger tau-associated atrophy rates (in magnitude) exhibit stronger FC degradation overall. This state-dependent coupling provides a parsimonious and biologically interpretable link between tau-driven tissue loss and systems-level functional disruption. Finally, while our analysis is cross-sectional, cognitive composites such as ADNI-MEM remain valuable endpoints for future work testing whether the inferred atrophy--FC coupling relates to memory decline at the individual level \citep{crane_development_2012}.

\subsection*{Limitations and Future Directions}
Several limitations delineate clear next steps. First, the limited availability of longitudinal fMRI in our dataset prevented within-subject modeling of FC trajectories; accordingly, our results and simulations are framed as a two-state (HC vs. AD) comparison. Second, our finite-element model assumes spatially uniform fiber density. In reality, regional differences in axonal packing and tract density would induce spatial heterogeneity in effective transport capacity, which is not captured here. And our current framework does not implement a closed-loop feedback in which severe axonal/synaptic loss would reduce the available routes for further propagation. Incorporating such structural feedback is a natural extension that could improve calibration in regions with pronounced tissue loss. Third, we presently focus on axonal-pathway-mediated spread and do not explicitly represent clearance mechanisms. A growing literature suggests that paravascular (glymphatic) exchange contributes to clearance of interstitial solutes, is modulated by arousal/sleep, and may influence amyloid-$\beta$ and tau dynamics \citep{iliff_paravascular_2012,xie_sleep_2013,dagum_glymphatic_2026}. Experimental work further implicates impaired glymphatic function in tauopathy models and provides mechanistic evidence that glymphatic pathways can clear extracellular tau \citep{harrison_impaired_2020,ishida_glymphatic_2022}. Incorporating multi-compartment transport and clearance terms—balancing axonal propagation with clearance—would therefore be a biologically motivated route to enhancing model realism.

\subsection*{Conclusion}
In summary, this study establishes a robust cross-sectional association between FC degradation and tau-linked atrophy parameters in AD and demonstrates that a simple atrophy-informed degeneration matrix can qualitatively reproduce the disease-state pattern of functional network disruption. By integrating multiphysics tau propagation and neurodegeneration with connectome-constrained whole-brain dynamics, the framework provides a mechanistically interpretable link from molecular pathology and tissue loss to systems-level functional breakdown. Beyond AD, the same strategy offers a general template for translating pathology-specific tissue degeneration into functional network impairment in related neurodegenerative proteinopathies.

\fi
\section*{Acknowledgements}
JK, JH and XW acknowledge the support from National Science Foundation (IIS-2011369) and National Institutes of Health (1R01NS135574-01). Data collection and sharing for this project was funded by the Alzheimer's Disease Neuroimaging Initiative (ADNI) (National Institutes of Health Grant U01 AG024904) and DOD ADNI (Department of Defense award number W81XWH-12-2-0012). ADNI is funded by the National Institute on Aging, the National Institute of Biomedical Imaging and Bioengineering, and through generous contributions from the following: AbbVie, Alzheimer's Association; Alzheimer's Drug Discovery Foundation; Araclon Biotech; BioClinica, Inc.; Biogen; Bristol-Myers Squibb Company; CereSpir, Inc.; Cogstate; Eisai Inc.; Elan Pharmaceuticals, Inc.; Eli Lilly and Company; EuroImmun; F. Hoffmann-La Roche Ltd and its affiliated company Genentech, Inc.; Fujirebio; GE Healthcare; IXICO Ltd.; Janssen Alzheimer Immunotherapy Research \& Development, LLC.; Johnson \& Johnson Pharmaceutical Research \& Development LLC.; Lumosity; Lundbeck; Merck \& Co., Inc.; Meso Scale Diagnostics, LLC.; NeuroRx Research; Neurotrack Technologies; Novartis Pharmaceuticals Corporation; Pfizer Inc.; Piramal Imaging; Servier; Takeda Pharmaceutical Company; and Transition Therapeutics. The Canadian Institutes of Health Research is providing funds to support ADNI clinical sites in Canada. Private sector contributions are facilitated by the Foundation for the National Institutes of Health (\href{http://www.fnih.org/}{www.fnih.org}). The grantee organization is the Northern California Institute for Research and Education, and the study is coordinated by the Alzheimer's Therapeutic Research Institute at the University of Southern California. ADNI data are disseminated by the Laboratory for Neuro Imaging at the University of Southern California. 

\section*{Authorship Contribution Statement}
JK: Methodology, Software, Validation, Investigation, Writing – Original Draft; JX: Methodology, Software, Validation; XW: Conceptualization, Validation, Supervision, Funding acquisition, Writing – Original Draft, Writing – Review and Editing.

\section*{Conflicts of Interest}
The authors declare that the research was conducted in the absence of any commercial or financial relationships that could be construed as a potential conflict of interest.

\selectlanguage{english}
\FloatBarrier%

\bibliographystyle{ieeetr} 
\bibliography{Bibliography/references,Bibliography/software,Bibliography/others}

\end{document}